\documentclass[pre,floatfix,a4paper,nofootinbib,amssymb,amsmath,showpacs,superscriptaddress]{revtex4}
\usepackage{graphicx}

\newcommand {\be} {\begin{equation}}
\newcommand {\ee} {\end{equation}}
\newcommand {\Be}{\begin{eqnarray*}}
\newcommand {\Ee} {\end{eqnarray*}}
\newcommand {\bey} {\begin{eqnarray}}
\newcommand {\eey} {\end{eqnarray}}
\newcommand{\bit}{\begin{itemize}}      
\newcommand{\eit}{\end{itemize}}
\newcommand{\bfl}{\begin{flusleft}}
\newcommand{\efl}{\end{flusleft}}
\newcommand{\bfr}{\begin{flushright}}
\newcommand{\ec}{\end{center}}
\newcommand{\ben}{\begin{enumerate}}    
\newcommand{\een}{\end{enumerate}}

\newcommand{\comment}[1]{}

\newcommand{\f}{\varphi}

\begin{document} 

\title{Stability of the splay state in pulse--coupled networks}

\author{R\"udiger Zillmer}
\email{ruediger.zillmer@gmail.com}
\affiliation{Istituto dei Sistemi Complessi, CNR,
CNR, via Madonna del Piano 10, I-50019 Sesto Fiorentino, Italy}
\affiliation{Lab. de neurophysique et de physiologie du systeme moteur,
CNRS-EP 1848  45 Rue des Saints-Peres,  75270 Paris, France}
\affiliation{INFN Sez. Firenze, via Sansone, 1 - I-50019 Sesto Fiorentino, Italy}
\author{Roberto Livi}
\email{livi@fi.infn.it}
\affiliation{Dipartimento di Fisica, Universit\'a di Firenze,
via Sansone, 1 - I-50019 Sesto Fiorentino, Italy}
\affiliation{Sezione INFN, Unita' INFM e Centro Interdipartimentale per lo Studio delle Dinamiche
Complesse, via Sansone, 1 - I-50019 Sesto Fiorentino, Italy}
\author{Antonio Politi}
\email{antonio.politi@isc.cnr.it}
\affiliation{Istituto dei Sistemi Complessi, CNR,
CNR, via Madonna del Piano 10, I-50019 Sesto Fiorentino, Italy}
\affiliation{Centro Interdipartimentale per lo Studio delle Dinamiche
Complesse, via Sansone, 1 - I-50019 Sesto Fiorentino, Italy}
\author{Alessandro Torcini}
\email{alessandro.torcini@isc.cnr.it}
\affiliation{Istituto dei Sistemi Complessi, CNR,
CNR, via Madonna del Piano 10, I-50019 Sesto Fiorentino, Italy}
\affiliation{INFN Sez. Firenze, via Sansone, 1 - I-50019 Sesto Fiorentino, Italy}
\affiliation{Centro Interdipartimentale per lo Studio delle Dinamiche
Complesse, via Sansone, 1 - I-50019 Sesto Fiorentino, Italy}

\begin{abstract}
The stability of the dynamical states characterized by a uniform firing rate
({\it splay states}) is analyzed in a network of globally coupled leaky
integrate-and-fire neurons. This is done by reducing the set of differential
equations to a map that is investigated in the limit of large network
size. We show that the stability of the splay state depends
crucially on the ratio between the pulse--width and the inter-spike interval.
More precisely, the spectrum of Floquet exponents turns out to consist of
three components: (i) one that coincides with the predictions of the mean-field
analysis [Abbott-van Vreesvijk, 1993]; (ii) a component measuring the
instability of ``finite-frequency" modes; (iii) a number of ``isolated" 
eigenvalues that are connected to the characteristics of the single pulse 
and may give rise
to strong instabilities (the Floquet exponent being proportional to the
network size). Finally, as a side result, we find that the splay state can be
stable even for inhibitory coupling.
\end{abstract}
   
\pacs{05.45.Xt,84.35.+i,87.19.La}

\maketitle


\section{Introduction}\label{one}

Understanding the mechanisms of information processing in the brain can be 
tackled by analyzing the dynamical properties of neural network models. 
Nonetheless, approaching the problem in full generality is a formidable task, 
since it requires taking into account, (i) the role of the topology of the 
connections, (ii) the dynamics of the connection themselves as a representation
of the synaptic plasticity, (iii) the internal dynamics of the model neurons,
which can depend on the number of ionic channels and other variables and 
parameters, (iv) the diversity among neurons and connections, (v) the 
unavoidable presence of noise. 
On the other hand, we can conjecture that at least some
basic mechanisms are quite robust and may depend on a few ingredients, only.
In fact, even simple models made of globally coupled identical 
units exhibit interesting dynamical properties,
far from being completely interpreted. For instance, the stability of 
their steady states is still a debated problem. 
In fact, one can find claims that the splay state \cite{nicols},
characterized by a uniform spiking rate of the neurons, is stable only in 
the presence of excitatory coupling \cite{abbott}, and yet stable splay 
states have been found also in networks with fully inhibitory coupling
\cite{zillmer}. The standard approach for 
determining the stability properties of such simple models
is based on the mean field approximation. This allows to obtain the
spectrum of eigenvalues associated with the stability matrix in the 
thermodynamic limit $N \to \infty$, $N$ denoting the number of neurons.
It is far from obvious if and how such results can be extended to the 
stability problem of large but finite networks. 
This is all more important if we consider that in the thermodynamic
limit the splay state has been shown to be marginally stable
for finite pulse--width, while it has been found to be strictly
stable for $\delta$-like pulses \cite{Zumdieck-Timme:2004,zillmer}. Moreover,
we have discovered that in the limit of vanishing pulse--width, the formula derived 
in \cite{abbott} does not coincide with the result of direct simulations.

We want to point out that assessing the linear stability of the splay
state is just a preliminary step towards a complete undertanding of the 
dynamical properties of neural networks. Nonetheless, even the 
accomplishment of this ``simple'' task is going to reveal unexpected
subtleties, which should be taken into account for pursuing further 
progresses. In particular, in this paper we introduce a specific 
formalism for determining the stability of the splay state in (in)finite
ensembles of neurons. The approach is not entirely new, as it is
based on the introduction of a Poincar\'e section, which transforms the 
original dynamical system into a map connecting dynamical configurations 
of the neural network after consecutive neural pulses (spikes)~.
A similar technique was used in \cite{Jin:2002} for obtaining 
analytic upper bounds of the maximum stability exponent of 
disordered neural networks with global inhibition. 
In \cite{bressloff} such a technique was exploited to 
setup a general approach for the investigation of various dynamical regimes, 
including the splay state. The analysis was performed by considering
the dynamical transformation describing consecutive spikes of the
same neuron: due to its complicacy such a transformation does not allow to
proceed much beyond formal statements. 
In this paper we have taken advantage from the idea of constructing a map 
between consecutive spikes of whatever neurons, combined with a suibtale shift 
of the neuron labels. In this way the computational complexity is significantly
reduced with respect to \cite{bressloff} and we are able to obtain analytic 
expressions also for large finite values of $N$.

One important consequence of our analysis is that one must
be extremely careful when dealing with the stability problem genereated by
finite pulse--width in large networks. Actually, even if the leading 
correction to the mean--field approximation is found to be ${\mathcal O}(1/N)$, 
we show that higher order corrections, up to ${\mathcal O}(1/N^3)$~, have to be 
taken into account for determining the correct stability properties,
even for arbitrarily large values of $N$.

Another important result of our study is that the thermodynamic limit does 
not commute with the zero pulse--width limit. In order to clarify this
issue we consider the stability problem in the presence of pulses, whose
width is inversely proportional to $N$. Although this recipe may appear a 
mathematical trick, it allows to point out a crucial aspect of the problem
at hand: when both $N$ and the spiking rate are large, their ratio is the only
relevant stability parameter. This implies also 
that the stability of a network subject to fixed small pulse-widths may 
qualitatively change when $N$ is increased, precisely because the 
above mentioned ratio varies.

In this paper we also clarify the basic question whether
a given network of pulse--coupled neurons exhibits a finite stability or if
it is ``marginally" stable. A general formula obtained in \cite{abbott} 
indicates that the amplitude of the eigenvalues of the stability spectrum
of the splay state converges to zero from below as $1/N^2$. This 
implies that infinitely large networks exhibit an arbitrarily large number of 
eigenvalues arbitrarily close to zero. Accordingly, one should conclude that 
marginal stability is a generic feature of infinite networks. This seems to 
contrast with the result of Ref.~\cite{zillmer}, where numerical evidence 
of stability in the thermodynamic limit was found for 
$\delta$-like pulses. Our analysis clarifies that the Abbott-Van Vreeswijk's 
formula \cite{abbott} applies to the case of finite pulse--width, while
in the case of vanishing pulse--width, stability is maintained also in the 
thermodynamic limit. The following argument provides 
a heuristic explanation of the mechanism generating such different stability
conditions. In networks of pulse--coupled
oscillators the same pulse acts differently on the various oscillators, because
they can be located in different regions of their phase space, when the
pulse is received. If the states of two neurons are close to each other in phase
space (e.g., the neurons exhibit almost equal values of their membrane potentials),
they will experience only slightly different effects from the same pulse
of finite width. This amounts to say that the two neurons are very weakly coupled,
thus yielding very small values of the Floquet stability exponent.
The same is no longer true when very small (infinitesimal) pulse--widths are considered, 
since the coupling field strongly oscillates also over ``microscopic'' time scales,
even for large values of $N$~. In other words, the ``roughness" of the coupling 
is able to strongly lock the neurons, thereby yielding a finite stability of the 
splay state.

By this argument one can also understand the limitations of the mean-field approach 
proposed in \cite{abbott}. In fact, once the neurons are ordered according to their
instantaneous potential, we have verified that in many cases the stability 
of the state is controlled by the stability of ``high-frequency" modes.
A typical example of such a mode is the perturbations associated with the 
forward (backward) shift of the even (odd) neurons (see Section IV).
These modes are by definition neglected in the mean--field 
approach \cite{abbott}.

In Sec. II we describe the model of leaky integrate-and-fire neurons 
considered in this paper and we reduce it in full generality 
to an event-driven map. Moreover, also the formalism for the linear stability
analysis is introduced. In Sec. III, we discuss the stability of the splay state 
in the case of finite pulse-width.
Sec. IV is devoted to the analsysis of vanishing ($1/N$) pulse-widths. 
A brief summary of the results and future perspectives are reported in Sec. V.


\section{The model}\label{two}

We consider a simple model of a network of identical 
leaky integrate-and-fire neurons, whose interaction is mediated by
a field $E(t)$~. This field can be viewed as the result of the linear
superposition of single--neuron pulses $E_s(t)$, emitted when the
membrane potential reaches a given threshold value. 
In agreement with Ref.~\cite{abbott}, we assume that the time dependence
of a single pulse is given by the relation
$E_s(t)= \alpha^2 t e^{-\alpha t}\,$, where $1/\alpha$ is the pulse--width.
Accordingly, in between consecutive spikes, the field evolution is
described by the differential equation,

\begin{equation}\label{eq:E}
  \ddot E(t) +2\alpha\dot E(t)+\alpha^2 E(t)=0\ .
\end{equation}

Moreover, the effect of the pulse emitted at time $t_0$ amounts to
a discontinuous change of the derivative,
\[
  \dot E(t_0^+)=\dot E(t_0^-)+\alpha^2/N\ .
\]
The dynamics of the membrane potential $x_i(t)$  of the $i$--th neuron 
is determined by the differential equation
\begin{equation}\label{eq:x1}
  \dot{x}_{i}=a-\eta x_{i}+gE(t)\, ,\ x_i\in (-\infty,1)\ ,\\
\end{equation}
where the parameter $g$ gauges the strength of the coupling with the
pulse field $E(t)$.
The membrane potential evolves according to 
(\ref{eq:x1}) until it reaches its threshold value $x_{i}=1\,$, when
it is instantaneously reset to the value $x_{i}=0$~. Such a condition
amounts to representing the discharge mechanism in neurons as a strong
nonlinear effect associated with a discontinuity in the membrane potential.
One should point out that this model is written from the very beginning 
in adimensional rescaled units (for a comparison with physical scales 
see the discussion in ref. \cite{zillmer}~)~.  
Without prejudice of generality one can further
simplify the equations by fixing the scale of time in units of the
parameter $\eta^{-1} = 1$. Consistently, one should formally rescale
also the parameter $\alpha$ to $\alpha/\eta$.

\subsection{Event-driven map}

By integrating Eq.~(\ref{eq:E}) between successive pulses, one can
reduce the continous--time evolution of the interaction field $E(t)$
to the discrete map,
\begin{subequations}\label{eq:map}
\begin{gather}
  E(n+1)=E(n) e^{-\alpha \tau}+NQ(n)\tau e^{-\alpha \tau}\\
  Q(n+1)=Q(n)e^{-\alpha \tau}+\frac{\alpha^2}{N^2}\ ,
\end{gather}
\end{subequations}
where $\tau$ is the interspike time interval
and we have introduced the new variable $Q := (\alpha E+\dot E)/N\,$~. 

Accordingly, also the differential equations (\ref{eq:x1}) can be 
integrated by taking into account this map-like representation
of the field evolution and reduced to an event--driven map 
for the membrane potential $x_i\,$,
\be\label{eq:xmap}
  x_{i}(n+1)=x_i(n)e^{-\tau}+a\left(1-e^{-\tau}
  \right)+gF(n)\ \qquad i=1,\dots,N \enskip.
\ee
where $F$ is
\begin{equation}\label{eq:F1}
  F(n)= \frac{{\rm e}^{-\tau} - e^{-\alpha\tau}}{\alpha-1}
     \left(E(n)+\frac{NQ(n)}{\alpha-1} \right) - 
  \frac{\tau e^{-\alpha\tau}}{(\alpha-1)} NQ(n)~.
\end{equation}

If the integer $m$ labels of the closest--to--threshold neuron, the interspike
time interval $\tau$ is obtained by imposing the condition, $x_m(n+1)=1\,$,
\begin{equation}\label{eq:ti}
  \tau(n)=\ln\left[\frac{x_m(n)-a}{1-gF(n)-a}\right]\ .
\end{equation}
Since in networks of identical neurons
the order of the potentials $x_i$ is preserved, it is convenient to introduce a
comoving spatial index frame, $j(n+1)=j(n)-1\,$, where the next firing neuron is
always labelled by $j(n)=1$ and reset to $j(n+1)=N\,$ after the firing event. 
Accordingly, the map for
the membrane potentials transforms into
\begin{equation}\label{eq:xx}
  x_{j-1}(n+1)=x_j(n)e^{-\tau}+1-x_1(n)e^{-\tau}\ \qquad j=1,\dots,N-1 \enskip,
\end{equation}
with the boundary condition $x_N=0$.
The set of Eqs.~(\ref{eq:map},\ref{eq:ti},\ref{eq:xx}) defines a discrete-time
mapping (notice that in the comoving frame the label $m$ in Eq.~(\ref{eq:ti})
is always set equal to 1)~, that is fully equivalent to the original set of 
ordinary differential equations. Altogether, it turns out that a network of 
$N$ identical neurons is described by $N+1$ equations: Two of them account for
the dynamics of $E(n)$, while the remaining $N-1$ equations describe the evolution
of the neurons ($x_N$ is no longer a variable being, by definition, equal to 0). 
Accordingly we see that, at variance with Ref.~\cite{bressloff}~ where no
field dynamics was explicitely introduced, the model has a finite 
dimension. 

In this framework, the periodic splay state reduces to a fixed point that
satisfies the following conditions,
\begin{subequations}\label{eq:ps}
\begin{gather}
  \tau(n)\equiv \frac{T}{N}\ ,\\
  E(n)\equiv\tilde E\, ,\ Q(n)\equiv\tilde Q\ ,\\
  \tilde x_{j-1}=\tilde x_j e^{-T/N}+1-\tilde x_1e^{-T/N}\ ,
\end{gather}
\end{subequations} 
where $T$ is the time elapsed between two consecutive spike emissions of
the same neuron. A simple calculation yields,
\[
  \tilde Q=\frac{\alpha^2}{N^2}\left(1-e^{-\alpha T/N}\right)^{-1}\,,\ 
  \tilde E=T\tilde Q\left(e^{\alpha T/N}-1\right)^{-1}\ .
\]
The solution of Eq.~(\ref{eq:ps}c) involves a geometric series that, together
with the boundary condition $\tilde x_N=0\,$, leads to a transcendental equation
for the period $T\,$. Not surprisingly, the result turns out to be independent
of the pulse--width $\alpha\,$. For simplicity, we report only the leading
terms for $N\gg 1\,$,
\begin{subequations}\label{eq:perxt}
\begin{gather}
  \tilde x_{N-j}=\frac{aT+g}{T}\left(1-e^{-j\,T/N}\right)\ ,\\
  T=\ln\left[\frac{aT+g}{(a-1)T+g}\right]\ .
\end{gather}
\end{subequations}
By assuming that the uncoupled neurons are in the repetitive firing regime,
i.e. by taking $a > 1$, the period $T$ is well defined in the excitatory
case ($g > 0$) only for $g < 1$ ($T\to 0$, when $g$ approaches 1), while in 
the inhibitory case ($g<0$), a meaningful solution exists for any coupling 
strength ($T\to \infty$ for $g \to -\infty$).

\subsection{Linear stability}

The main complication of the linear stability analysis stems from the 
implicit dependence of $\tau$ on the relevant variables (see Eq.~\eqref{eq:ti},
where $m=1$ in the comoving frame we have adopted). By linearizing
Eq.~\eqref{eq:ti}, we obtain
\[
  \delta \tau(n) =\tau_x \delta x_1(n) +\tau_E\delta E(n)+\tau_Q\delta Q(n)\ ;
\]
where $\tau_x:=\partial \tau/\partial x_1$ and analogous definitions hold for
$\tau_E$ and $\tau_Q$. The linear stability of the splay state is thus 
determined by the dynamics of the perturbations $\delta x_j$, $\delta E$, and
$\delta Q$. This is expressed by the following set of linear equations
\begin{subequations}\label{eq:lin1}
\begin{gather}
  \delta x_{j-1}(n+1)=e^{-T/N}[\delta x_j(n)-\delta x_1(n)]+ e^{-T/N}
  (\tilde x_1-\tilde x_j)\delta \tau(n)\,,\\
  \delta E(n+1)=e^{-\alpha T/N}\delta E(n)+Te^{-\alpha T/N}\delta Q(n)
  -\left(\alpha\tilde E-N\tilde Qe^{-\alpha T/N}\right)\delta \tau(n)\,,\\
  \delta Q(n+1)=e^{-\alpha T/N}\delta Q(n)-\alpha\tilde Qe^{-\alpha T/N}
  \delta \tau(n)\, ,
\end{gather}
\end{subequations}
which have been linearized around the fixed point (\ref{eq:ps})~.
The boundary conditon $x_N\equiv 0\,$ imposed by the comoving frame yields 
$\delta x_N=0\,$. In practice, the stability problem is solved by computing
the Floquet spectrum of multipliers $\{\mu_k\}\,$, $k=1,\ldots,N+1\,$ 
associated with the eigenvalue problem of the 
set of linear equations \eqref{eq:lin1}~. In general the explicit computation
has to be performed numerically.

For the sake of clarity, it is first convenient to discuss the trivial
case of vanishing interaction, i.e. $g=0\,$. After simple calculations
the eigenvalues turn out to be $\mu_k=\exp(i\varphi_k)$, where
$\varphi_k= \frac{2\pi k}{N}$, $k=1,\ldots,N-1$, and $\mu_{N}=\mu_{N+1}=
\exp(-\alpha T/N)\,$. The last two exponents concern the dynamics of
the coupling field $E(t)$, whose decay is ruled by the time scale $\alpha^{-1}\,$.

\begin{figure}[t!]
\includegraphics[draft=false,clip=true,height=0.34\textwidth]{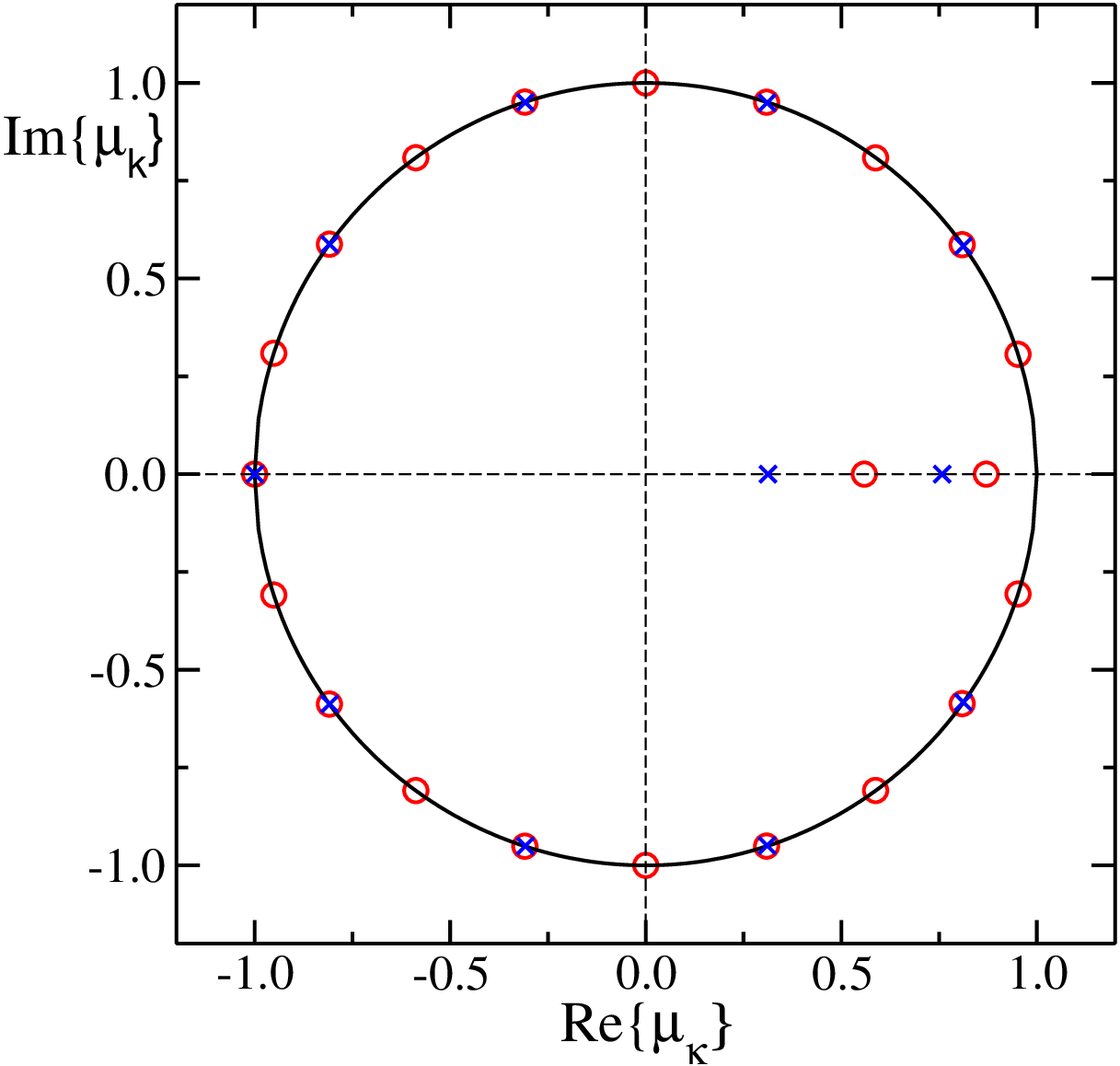}
\caption{
(Color online) Unitary circle and exact Floquet multipliers spectra for the 
complete maps for $N=20$ (red circles) and $N=10$ (blue crosses) for
excitatory coupling. The parameters are $a= 3.0$, $g=0.4$, and $\alpha=30.0$.
}
\label{fig1}
\end{figure}

As soon as the coupling is on, small amplitude fluctuations 
$ \sim {\mathcal O}(g/N)$ 
affect the neuron dynamics and the spectrum of Floquet multipliers 
takes the general form   
\begin{gather}\label{eq:specdef}   
  \mu_k=e^{i\varphi_k}e^{T(\lambda_k+i\omega_k)/N}\,,\ \varphi_k=
  \frac{2\pi k}{N}\,,\ k=1,\ldots,N-1\,,\\
  \mu_{N}=e^{T(\lambda_{N}+i\omega_{N})/N}\,,\
  \mu_{N+1}=e^{T(\lambda_{N+1}+i\omega_{N+1})/N}\,.\nonumber
\end{gather}
where, $\lambda_k$ and $\omega_k$ are the real and imaginary parts of the
Floquet exponents. 
As an example, in Fig.~\ref{fig1} we show the spectrum of the Floquet
multipliers of the splay state for excitatory coupling ($g > 0$)~ and finite
values of $N$~. The multipliers with $k=1,\ldots,N-1$ are very close
to the unit circle, while the two isolated multipliers $\mu_{N}$ and
$\mu_{N+1}$ lay very close to the real axis inside the unit circle.
We want to point out that already for $g/N \approx {\mathcal O}(10^{-2})$
the multipliers of the coupled case can be viewed as small ``perturbations"
of the uncoupled one. This observation supports the perturbative approach
that we are going to discuss in the following sections.

Before accomplishing this task, it remains to comment that
the variable $\varphi_k$ plays the same role of the
wavenumber in the linear stability analysis of spatially extended systems, 
so that we can say that $\lambda_k$ characterizes the stability of the $k$--th
mode. In the following analysis it is convenient to distinguish between
modes characterized by $\varphi_k \approx 0, \mod (2\pi) + {\mathcal O}(1/N)$, 
and the other modes. Actually they identify two spectral components, that require 
a different mathematical treatment. The first component corresponds
to the condition $||\mu_k-1|| \sim N^{-1}$ and is referred to as 
{\it long wavelengths} (LWs), while the second one corresponds to 
$||\mu_k-1|| \sim {\mathcal O}(1)$ and is referred to as, {\it short wavelengths} (SWs).

\section{Finite pulse--width}\label{sec:fsw}

In this section we investigate the stability problem of the splay state
for networks subject to pulses with finite $\alpha$ (i.e. independent of 
the system-size) for large values of $N$.
This is also the setup studied in \cite{abbott} by a mean field analysis.
By retaining all the terms up to the order $1/N$, 
the event--driven map (see Eqs.(\ref{eq:map}) and (\ref{eq:xmap})~)
simplifies to the set of $N+1$ equations
\begin{subequations}\label{eq:mapa}
\begin{gather}
  E(n+1)=(1-\alpha \tau)E(n)+NQ(n)\tau\ ,\\
  Q(n+1)=(1-\alpha \tau)Q(n)+\frac{\alpha^2}{N^2}\ ,\\
  x_{j-1}(n+1)=(1-\tau)x_j(n)+1-x_1(n)+\tau\ ,
\end{gather}
\end{subequations} 
where $j=1,\ldots,N-1$. Here the expression of interspike interval
(\ref{eq:ti}) simplifies to
\begin{equation}\label{eq:tia}
  \tau(n)=\frac{1-x_1(n)}{a-1+gE(n)}\ .
\end{equation}
The periodic solution for the pulse--field becomes $\tilde E=T^{-1}\,$, and $\tilde Q
=\alpha/NT\,$, while $\tilde x_j$ and the period $T$ are still given by 
Eq.~\eqref{eq:perxt}. The Floquet eigenvalue spectrum $\mu_k\,$, $k=1,\ldots,N+1\,$,
can be obtained by solving the set of $N+1$ linear equations
\begin{subequations}\label{eq:lina}
\begin{eqnarray}
  \mu_k\delta x_{j-1} &=& (1-T/N)\delta x_j-\delta x_1+ (1-\tilde x_j)
  \delta \tau\,,\\
  \mu_k\delta E &=& (1-\alpha T/N)\delta E+T\delta Q\,,\\
  \mu_k\delta Q &=& (1-\alpha T/N)\delta Q-\frac{\alpha^2}{NT}\delta \tau\,.
\end{eqnarray}
\end{subequations}
An explicit expression for $\delta \tau$ is obtained by evaluating the
derivatives of Eq.~\eqref{eq:tia} 
\[
  \delta \tau=-\frac{T^2}{N}\,\delta E-\frac{T}{g}\,\delta x_1\ .
\]
With this substitution in Eqs.(\ref{eq:lina}) we conclude that the eigenvalue
problem amounts to finding the $N+1$ roots $\mu_k$ of the associated
polynomial. 
A partial simplification of the problem can be obtained by 
extracting from Eqs.(\ref{eq:lina} b,c) the dependence of 
$\delta \tau$ directly in terms of the perturbation of the 
potential of the next--to--threshold neuron $\delta x_1$,
\[
  \delta \tau=K\delta x_1\,
\]
where
\begin{equation}
  K=\left[-(a-1+g/T)+\frac{\alpha^2gT}{N^2(\mu_k-1+\alpha T/N)^2}\right]^{-1}
  \label{eq:K1}\,.
\end{equation}
Finally, by substituting this expression into into Eq.~(\ref{eq:lina}a), we obtain 
a closed set of equations for the perturbations of the membrane 
potentials,
\begin{equation}\label{eq:linxa}
  \mu_k\delta x_{j-1}=(1-T/N)\delta x_j+( K-1)\delta x_1- K
  \tilde x_j\delta x_1\ .
\end{equation}
After imposing the boundary condition, $\delta x_N=0\,$, Eq.~\eqref{eq:linxa}
reduces to the following eigenvalue equation,
\begin{equation}\label{eq:speca1}
  \mu_k^{N-1}e^{T}=K(a+g/T)\frac{1-\mu_k^{N-1}}{1-\mu_k}
  -[K(a-1+g/T)+1]\frac{1-\mu_k^{N-1}e^{T}}{1-\mu_ke^{T/N}}\ ,
\end{equation}
where $K$ is a function of $\mu_k\,$ (see Eq.~\eqref{eq:K1})~.

In order to solve analytically the above equation, it is necessary to distinguish 
between long and short wavelengths.

\subsection{Long wavelengths}\label{sec:ecua}

Let us consider the modes for which  
$||\mu_k-1||\sim N^{-1}\,$, or, equivalently, 
$\varphi_k \approx 0, \mod (2\pi) + {\mathcal O}(1/N)$~.
In order to simplify the notation we define
\[
  \Lambda_k:=\frac{N}{T}\ln \mu_k\,.
\]
At leading order, $K$ writes
\[
  K=\left[-(a-1+g/T)+\frac{\alpha^2g}{T(\Lambda_k+\alpha)^2}\right]^{-1}\ .
\]
Moreover, making use of the relation $\tilde E=T^{-1}$, the eigenvalue 
equation \eqref{eq:speca1} can be approximated as
\begin{equation}\label{eq:speca2}
  \frac{a T+g}{\alpha^2g}\left(1-e^{-\Lambda_k T}\right)
  (\Lambda_k+\alpha)^2(\Lambda_k+1)=\Lambda_k\left(
  e^{T}-e^{-\Lambda_k T}\right)\,,\ ||\Lambda_k||\ne 0\,.
\end{equation}
This equation coincides with that one derived from the mean field analysis
\cite{abbott}, except for the missing $||\Lambda_k|| = 0\,$, which
corresponds to the trivial zero Floquet exponent of the continuous time 
evolution \eqref{eq:x1} and disappears in the discrete-time dynamics.
We want to stress again that, despite Eq.~(\ref{eq:speca2}) yields 
$N+1$ roots, it provides a suitable approximation only for those
eigenvalues which satisy the relation $||\mu_k-1||\sim N^{-1}\,$.

\subsection{Short wavelengths}

The second component of the spectrum \eqref{eq:specdef} is obtained for 
$||\mu_k-1|| \sim {\mathcal O}(1)$. In this case $K$ has the 
simple form
\[
  K=-(a-1+g/T)^{-1}\ .
\]
Upon introducing into Eq.~\eqref{eq:speca1} the explicit form of the periodic
solution (\ref{eq:perxt}b), the spectrum simplifies considerably, namely
\begin{equation}\label{eq:speca3}
  \mu_k^N=e^{-T}\frac{a+g/T}{a-1+g/T}=1\ ,
\end{equation}
i.e., it coincides with a fully degenerate Floquet spectrum,
\begin{equation}\label{eq:lspeca}
  \omega_k\equiv 0\,,\ \lambda_k\equiv 0\,
\end{equation}
Notice that this approximation holds only for those eigenvalues such that
$||\mu_k-1|| \sim {\mathcal O}(1)$, i.e. those representing 
the large maiority of the spectrum, except those laying close to the point (1,0), 
where the unit circle intercepts the real axis (see Fig.~\ref{fig1})\, .

For what concerns the isolated eigenvalues $\mu_N$ and $\mu_{N+1}$
one can easily argue that for finite $N$ and finite pulse--widths
they are always contained inside the unit
circle close to the real axis  and they approch the point $(1,0)$ 
from the left as $N\to \infty$\, . Accordingly, they can at most
contribute to the marginal stability of the dynamics. 

\subsection{Phase diagram and finite-size corrections}

\begin{figure}[t!]
\includegraphics[draft=false,clip=true,height=0.34\textwidth]{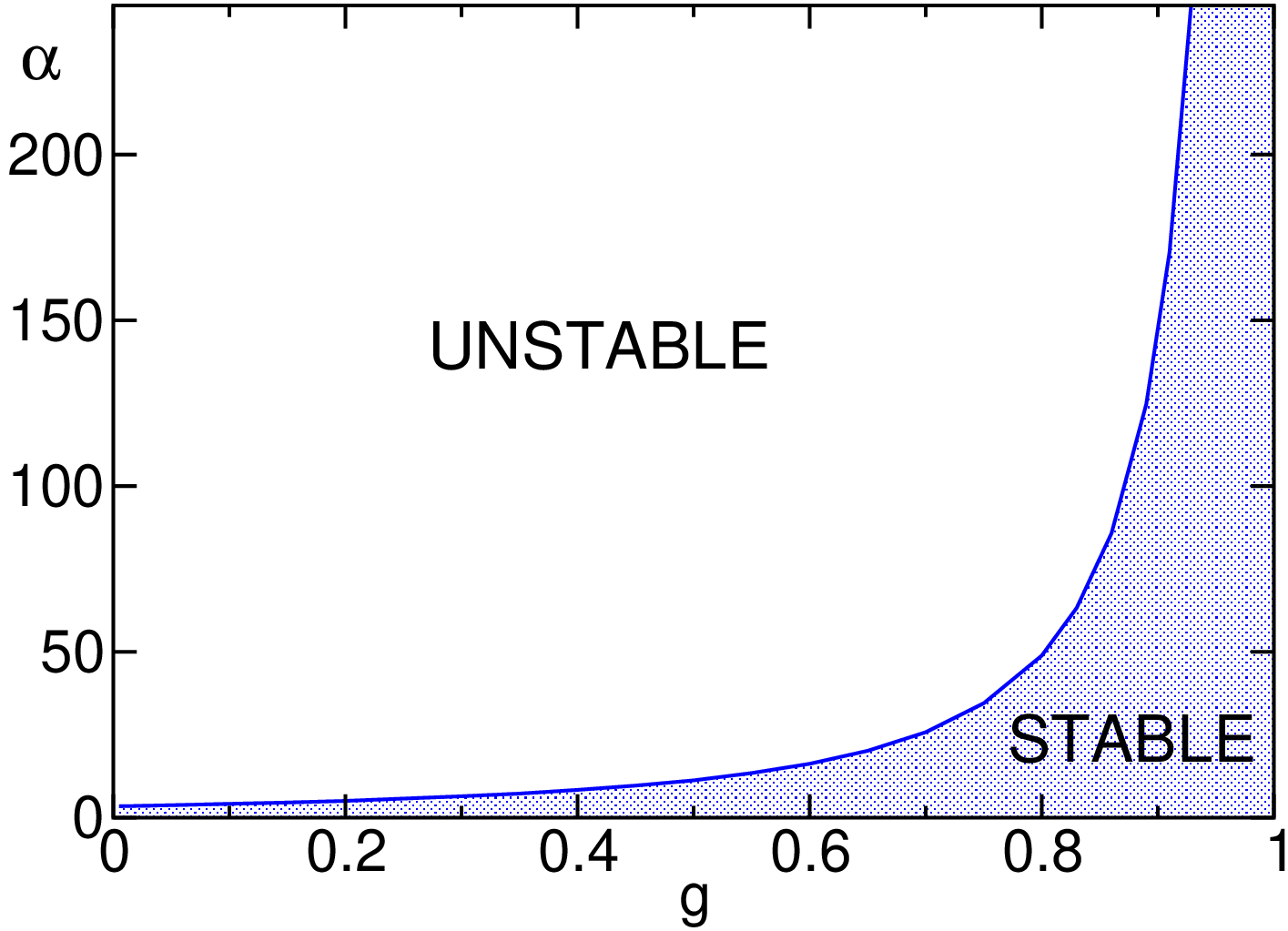}
\caption{Phase diagram for the stability of the splay state in a 
neural network with excitatory coupling acting through finite
pulse--width. The solid line separating the stable
from the unstable regions in the $(g,\alpha)$-plane has been
derived from the analytic formula of the Floquet spectrum (\ref{eq:speca2})
with $a=1.3\,$. Please notice that in this context {\it stable} refers to
finite $N$, for infinite systems the stability will become marginal.
}
\label{fig2}
\end{figure}

According to the above results one can conclude that in the
limit $N \to \infty$ the onset of instabilities is determined by the Floquet
exponents associated with LCs (see Eq.~\eqref{eq:speca2}). 
It is, therefore, not surprising to discover that our result coincides with 
the predictions of the mean--field analysis reported in \cite{abbott}. We
also confirm that the splay state is always unstable for inhibitory coupling
($ g < 0 \,$). In the case of excitatory coupling the
mean--field analysis predicts stability of the the splay state
for $\alpha \le \alpha_c(g,a)\,$, where the $\alpha_c(g,a)$ is the
critical line separating the stable from the unstable region (see 
Fig.~\ref{fig2}))\,. By including the role of SWs
we can conclude that in the limit $N \to \infty$ the splay state can
be at most marginally stable for $\alpha \le \alpha_c(g,a)\,$.
In particular, $\alpha_c$ diverges to $+ \infty$ for $g \to 1$ (see Fig.~\ref{fig2}). 
Our analysis confirms also that the splay state becomes always 
unstable via a Hopf-bifurcation, that gives rise to collective periodic
states \cite{vvres,mohanty}. For $g > 1$ no splay state can exist
and this can be viewed as a drawback of the 
model: for too strong excitatory coupling, no stationary regime can be
sustained, since the evolution steadily accelerates.

\begin{figure}[t!]
\includegraphics[draft=false,clip=true,height=0.34\textwidth]{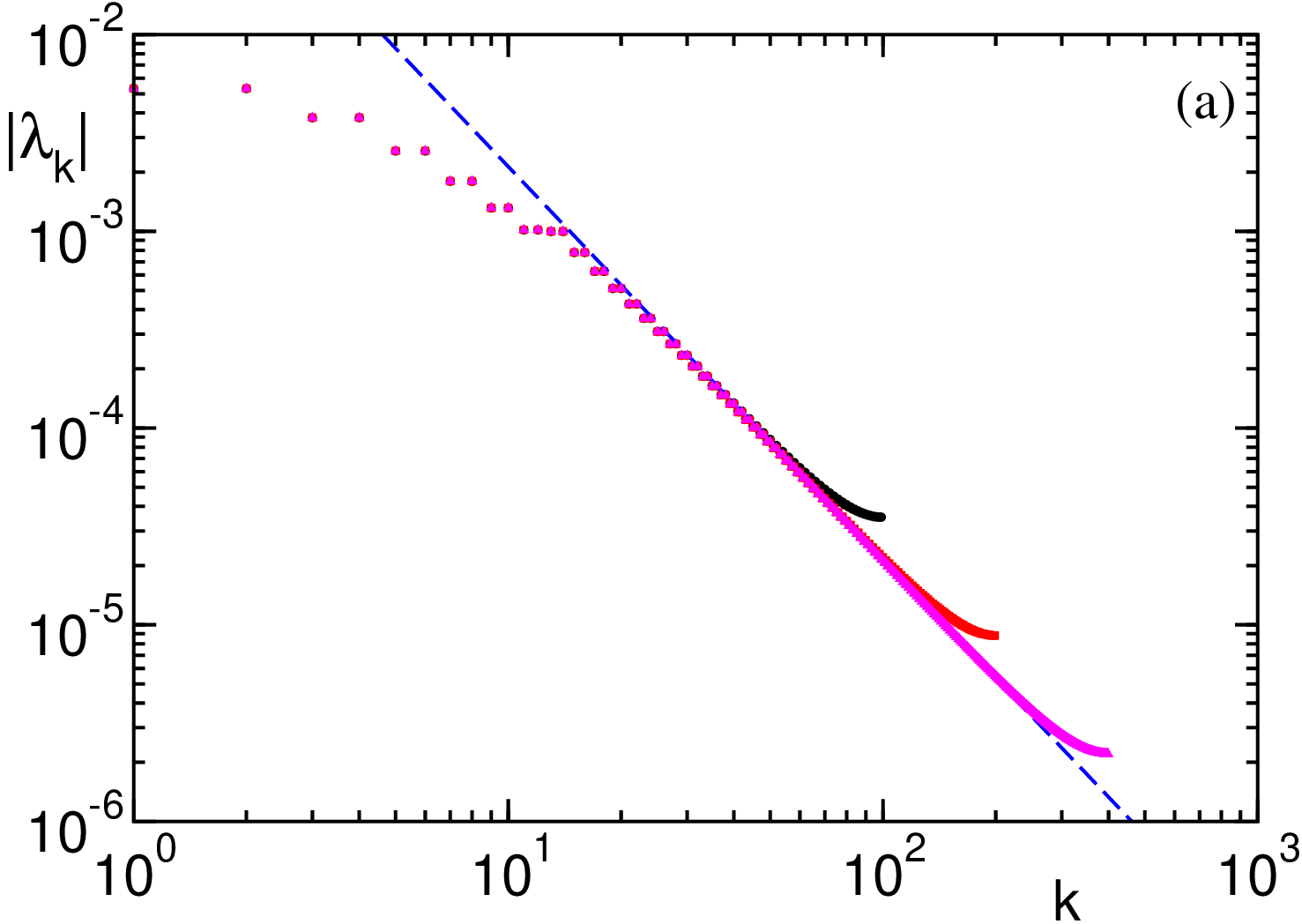}
\includegraphics[draft=false,clip=true,height=0.34\textwidth]{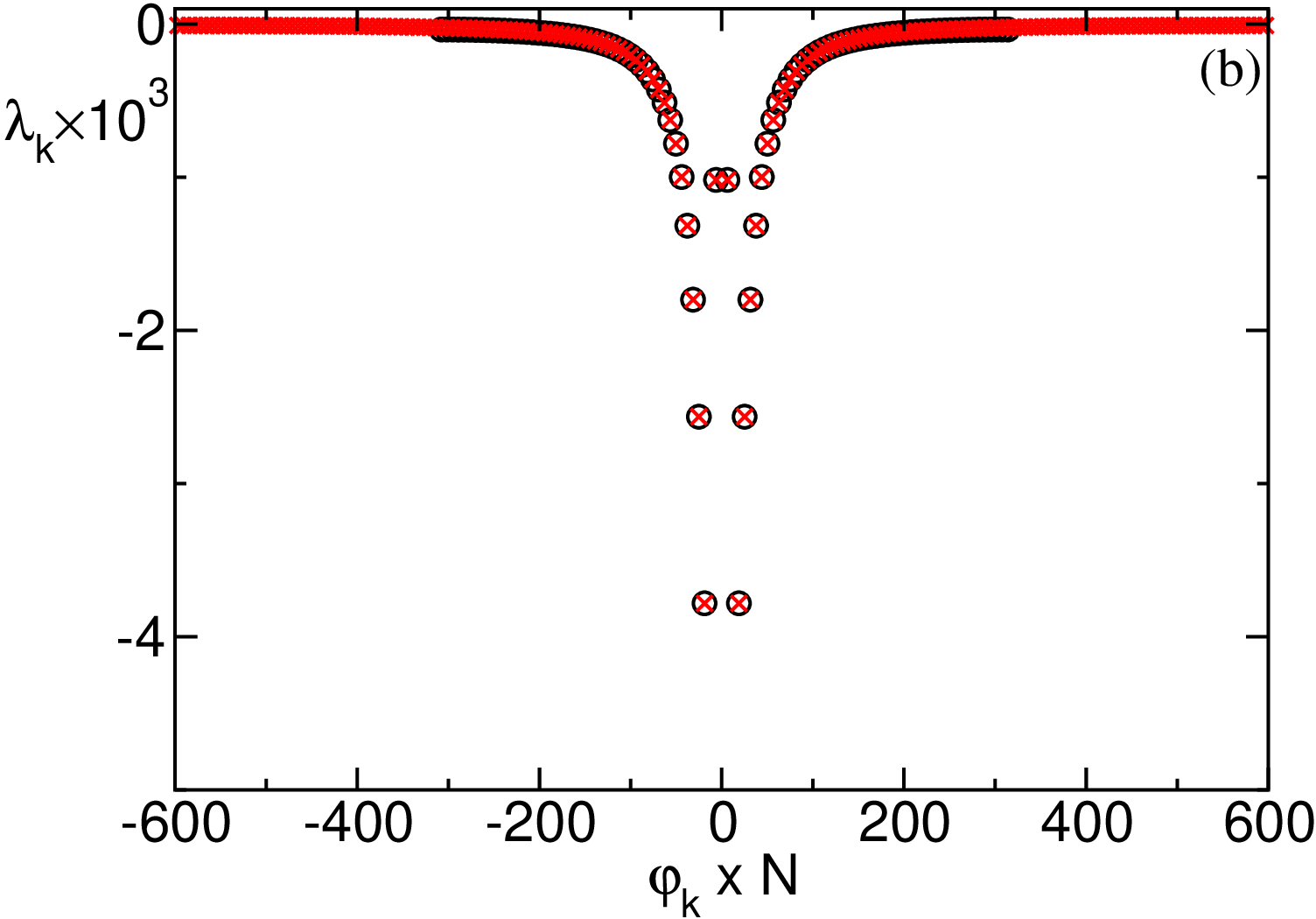}
\caption{
(Color online) (a) Log-log plot of the absolute values of the the Floquet 
exponents $\lambda_k$, ordered from the largest to the smallest as a 
function of the index $k=1, ..., N$ for $N=100, 200, 400$. The dashed line 
has slope -2. 
(b) The Floquet exponent as a function of the rescaled phase $\f N$, for 
$N=100$ (black circles) and $N=200$ (red crosses). In both pictures the 
parameter values are $a= 3.0$, $g=0.4$, and $\alpha=30.0$.}
\label{fig3}
\end{figure}

\begin{figure}[t]
\includegraphics[clip=true,height=0.34\textwidth]{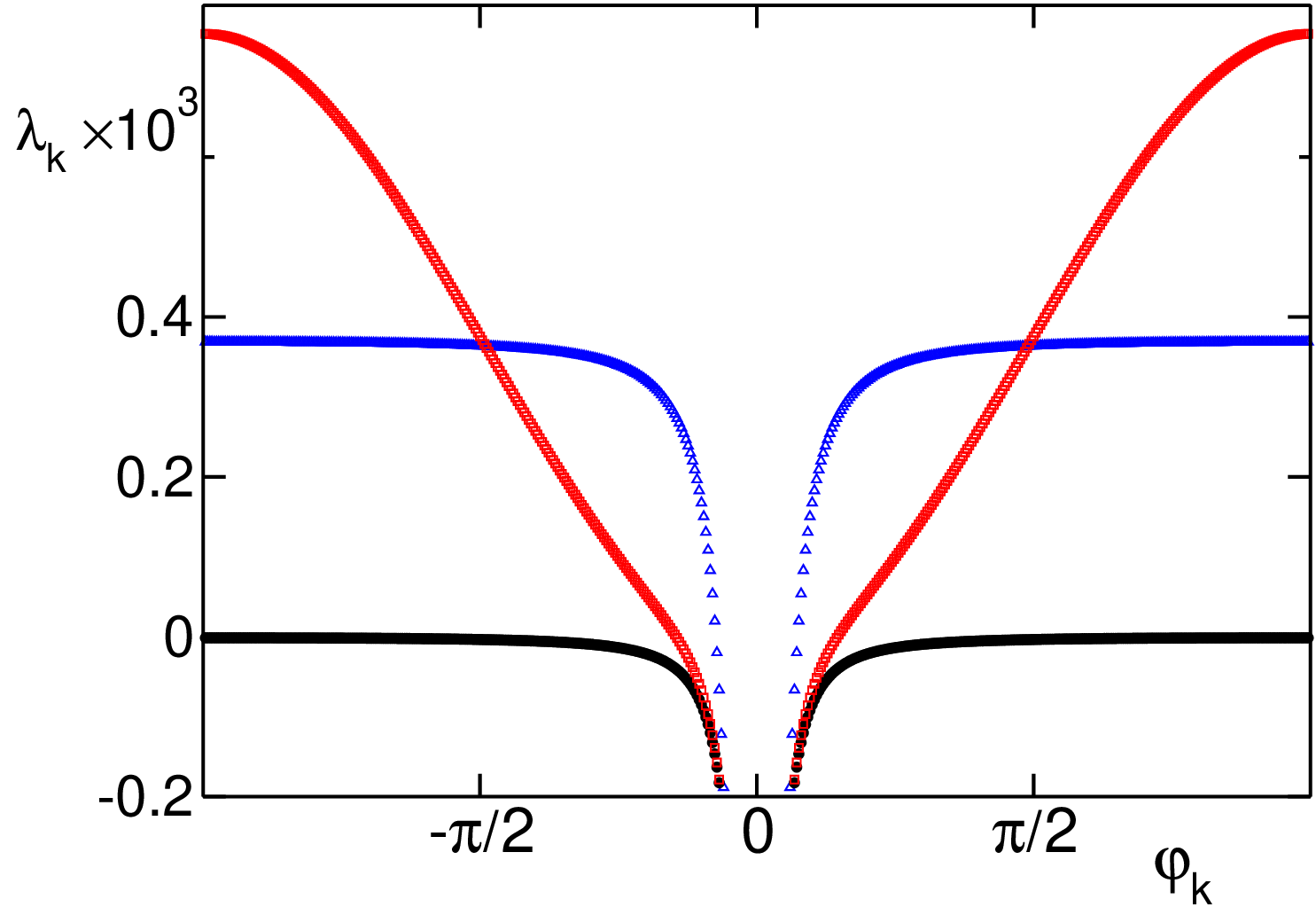}
\caption{
(Color online) Floquet exponents $\lambda_k(\f)$ as a function of the phase $\f_k$
for finite pulse--width $\alpha=3.0$ and finite neuron numbers $N=500$ in the case
of excitatory coupling $g=0.4$. The filled
circles represent the exact result for finite $N$, while the red empty squares and
the blue empty triangles refer to approximated results correct up to the first
or second order in $1/N$, respectively.
The parameters here are $a= 0.3$, $\eta=0.1$.
}
\label{fig4}
\end{figure}

The perfect degeneracy of the zero Floquet exponents
associated to SWs sheds doubts on the effective stability properties 
of large but finite networks, since such modes turn out to be marginally stable. 
Therefore, we have
decided to solve numerically the eigenvalue equations (\ref{eq:lin1}) for 
different system sizes and for values of the parameters $g$ and $\alpha$,
which correspond to marginally stable splay states in the limit 
$N \to \infty$. The results plotted in Fig.~\ref{fig3}) show that the 
splay state is strictly stable in finite lattices and that the maximum Floquet 
exponent approaches zero from below as $1/N^2$. This implies 
that a $1/N$ perturbation theory, like the one developed in this section, 
cannot account for such deviations and even a second-order approximation scheme
cannot capture the instabilities of the original model. This is confirmed in
Fig.~\ref{fig4}, where the Floquet spectra obtained from first- and
second-order approximations are seen to yield an unstable splay state, 
even though the numerical solution of the stability problem indicates that
the finite--$N$ model is stable.

As the event-driven map is the standard approach adopted to simulate
this type of networks, an important consequence of our analysis is that
one should be careful enough to consider at least third order approximations
in the $1/N$ perturbative expansion, otherwise the resulting approximate equations 
may fail to reproduce the correct asymptotic dynamics.

\section{Vanishing pulse--width}

In \cite{zillmer}, we have investigated the synchronization properties of a
network of leaky integrate-and-fire neurons in the case of $\delta$-like pulses.
One might expect that the stability properties of such a network
can be determined by taking the limit of vanishing pulse--width in the 
formulae obtained in Ref.~\cite{abbott}. However, the lack of agreement 
with numerics suggests that the $N\to \infty$ limit does not commute with 
the zero pulse--width limit. In order to clarify this issue, we introduce a 
new setup, where the pulse-width is rescaled with the network size $N$ as
$N^{-1}\,$. This amounts to impose a decay rate proportional to $N$
\[
  \alpha :=  \beta N .
\]
It is important to notice that in this context we have to deal with two time scales:
(i) a scale of order ${\mathcal O}(1)$ that corresponds to the evolution of the
the membrane potential $x_j$; (ii) a scale of order $\alpha^{-1}\sim N^{-1}\,$
that corresponds to the field relaxation.

At leading order, the expression \eqref{eq:F1} for the auxiliary
quantity $F(n)$ simplifies to
\[
  F(n)=\frac{1}{N\beta^2}\left[\beta E(n)\left(1-e^{-N\beta\tau}\right)
  +Q(n)\left(1-e^{-N\beta\tau}-N\beta\tau\,e^{-N\beta\tau}\right)\right]\ .
\]
The splay state corresponds to the fixed--point 
\[
  \tilde Q=\beta^2\left(1-e^{-\beta T}\right)^{-1}\,,\ 
  \tilde E=T\tilde Q\left(e^{\beta T}-1\right)^{-1}\ ,
\]
while the derivatives of the interspike interval \eqref{eq:ti} now read
\[
  \tau_x=\frac{N}{1-a-g\tilde E}\,,\ \tau_E=\frac{g}{N\beta}\left(1-e^{-\beta T}\right)
  \tau_x\,,\ \tau_Q=\frac{g}{N\beta^2}\left(1-e^{-\beta T}-\beta Te^{-\beta T}\right)\tau_x\ .
\]
Finally, the eigenvalue spectrum is defined by the set of linear equations
\begin{subequations}\label{eq:evb}
\begin{eqnarray}
  \mu_k\delta x_{j-1} &=& (1-TN^{-1})\delta x_j-\delta x_1
  + N^{-1}(1-\tilde x_j)\delta \tau\,,\\
  \mu_k\delta E &=& e^{-\beta T}\delta E+Te^{-\beta T}\delta Q
  -N\left(\beta\tilde E-\tilde Qe^{-\beta T}\right)\delta \tau\,,\\
  \mu_k\delta Q &=& e^{-\beta T}\delta Q-N\beta\tilde Qe^{-\beta T}
  \delta \tau\,.
\end{eqnarray}
\end{subequations}
We repeat the same formal analysis illustrated in Sec.~\ref{sec:fsw}, by first 
solving the eigenvalue equations for the field perturbations $\delta x_j$\, . 
After some tedious calculations we obtain $\delta \tau=K\delta x_1\,$, where
\begin{equation}\label{eq:bk1}
  K=\frac{(\mu_ke^{\beta T}-1)^2}{(\mu_ke^{\beta T}-1)^2(1-a-g\tilde E)+
  \mu_k\beta^2Tg\,e^{\beta T}}\ .
\end{equation}
By substituting this expression into Eq.~\eqref{eq:evb} and by imposing
the boundary condition $\delta x_N=0$, we obtain a closed equation for the
spectrum $\{\mu_k\}\,$, $k=1,\ldots,N+1\,$,
\begin{equation}\label{eq:specb1}
  \mu_k^{N-1}e^{T}= KC\frac{1-\mu_k^{N-1}}{1-\mu_k}
  -[ K(C-1)+1]\frac{1-\mu_k^{N-1}e^{T}}{1-\mu_ke^{T/N}}\ ;
\end{equation}
where $C:=(g+aT)/ T\,$ and $K$ is still a function of $\mu_k$.
Proceeding along the previous guidelines, it is necessary to separately
discuss the behaviour of long and short wavelengths.

\subsection{Long wavelengths}

As well as in Sec.~\ref{sec:ecua}, we consider the spectral
component characterized by $||\mu_k-1|| \sim N^{-1}\,$. 
In this case $K$ assumes the simple expression 
\[
  K=-\frac{1}{a-1} .
\]
By neglecting $1/N$ terms, the spectrum satisfies the equation,
\begin{equation}\label{eq:specb2}
  \left(1-e^{T\Lambda_k}\right)(1+ \Lambda_k^{-1})=\left(1-
  e^{T(\Lambda_k+ 1)}\right)\frac{g}{aT+g}\ ,
\end{equation}
where $\Lambda_k:=NT^{-1}\ln \mu_k\,$. 
This result agrees again with the mean field analysis \cite{abbott} in the limit
$\alpha\sim N\gg 1\,$. Let us remark that, despite Eq.~(\ref{eq:specb2})
yields $N+1$ eigenvalues, it provides a good approximation only for those
such that $||\mu_k-1|| \sim N^{-1}\,$. 

\subsection{Short wavelengths}

For $||\mu_k-1|| \sim {\mathcal O}(1)$ we can replace the term
$(\mu_ke^{\beta T}-1)$ with $(e^{i\varphi_k+\beta T}-1)$, in
Eq.~(\ref{eq:bk1}), thus obtaining $K=K(\varphi_k)\,$, for 
$k=1,\ldots,N-1\,$. Moreover, by using the
expression (\ref{eq:perxt}b) for the period $T$, one can simplify
Eq.~\eqref{eq:specb1} to the following form
\begin{equation}\label{eq:specb3}
  \mu_k^N=e^{-T}(1- K)\,.
\end{equation}
Accordingly, the Floquet spectrum writes 
\begin{equation}\label{eq:lspecb3}
  \lambda_k+i\omega_k=-1+\frac{1}{T}\ln[1- 
  K(\varphi_k)]-\frac{i\varphi_k N}{T}\,.
\end{equation}
For even values of $N$ and for the maximal phase $\varphi_{1 + N/2}=\pi\,$,
an explicit solution for the Floquet exponent can be obtained:
\begin{equation}\label{eq:lanab}
  \lambda_{\pi} := \lambda_{1 + N/2}            =-1+\frac{1}{T}
  \ln\left[1+\frac{1}{a-1+2\beta^2Tg\left(1+e^{2\beta T}\right)
  \left(e^{3\beta T}-2e^{\beta T}+e^{-\beta T}\right)^{-1}}\right]\,.
\end{equation}

\subsection{Isolated eigenvalues}

The main consequence of the stability analysis discussed at
the beginning of this section is that there exist also eigenvalues whose 
time scale is of the order $\alpha^{-1}\sim N^{-1}$.
These exponents are those labelled with the indices $N$ and $N+1$
in Eq.~(\ref{eq:specdef}).  The analysis of
these two exponents becomes particularly relevant when they approach
and cross the unitary circle; their analytical expression can be easily 
derived by investigating the case $||\mu_{N,N+1} - 1||\sim 1\,$, which leads to 
a divergence of the maximal Floquet exponent, $\lambda_{m} \sim N\,$.
The two corresponding eigenvalues turn out to satisfy the equation
\begin{equation}\label{eq:mubeta}
\frac{\lambda_{N,N+1}}{N} =  -\beta +
\frac{1}{T}\ln\left | 1-\frac{\beta^2Tg}{2(1-a-g\tilde E)}\left(
  1 \pm \sqrt{1-\frac{4(1-a-g\tilde E)}{\beta^2Tg}}\right)\right |
\end{equation}
Whenever one of such solutions become positive, this will give a 
leading Floquet exponent $\lambda_{m} \sim N\,$.

\begin{figure}[t!]
\includegraphics[draft=false,clip=true,height=0.34\textwidth]{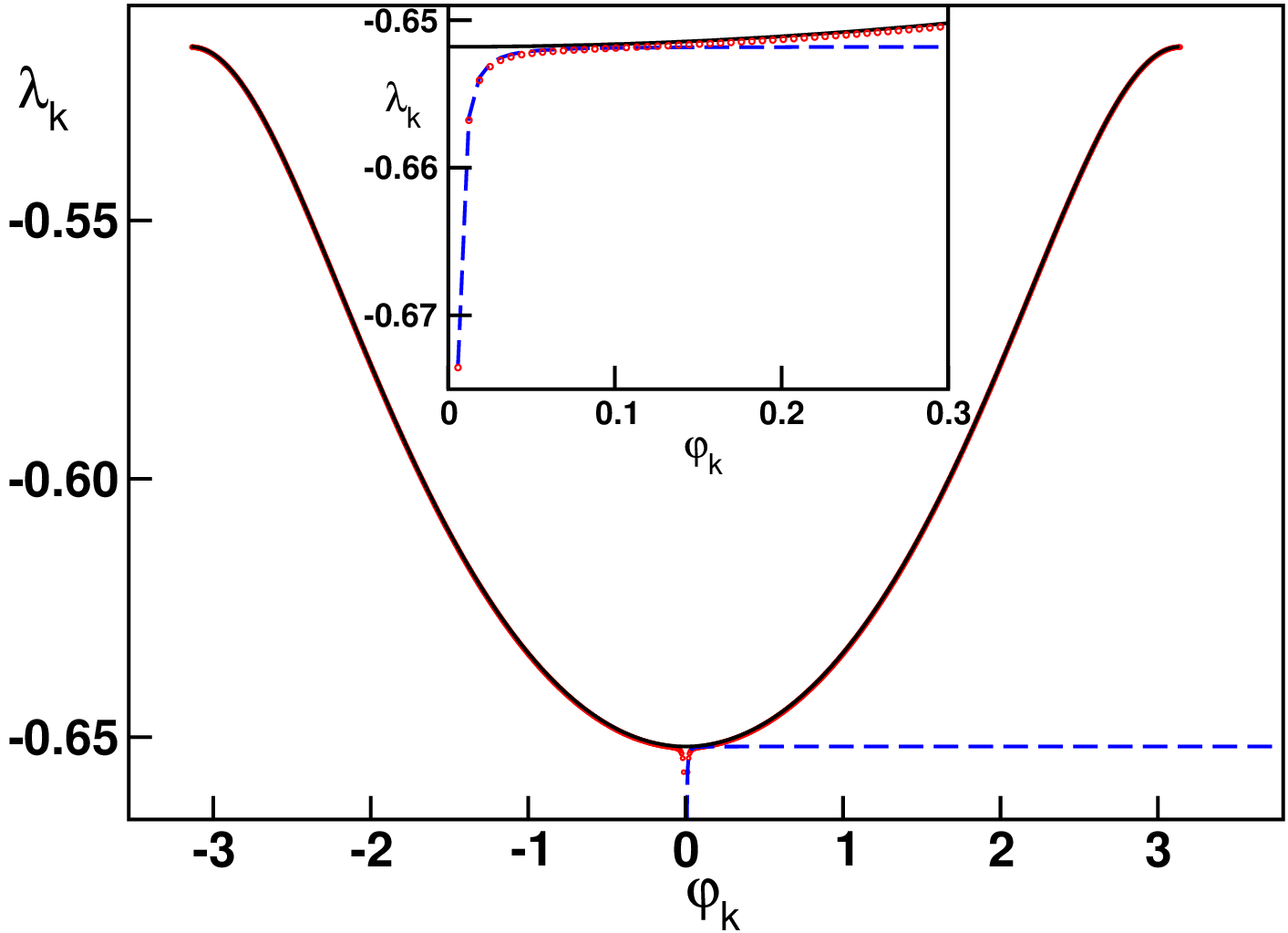}
\caption{
(Color online) Floquet spectrum $\lambda(\f)$ as a function of the phase $\f$
in the case of vanishing pulse--width. Red circles indicate the numerical results
obtained for the linear stability analysis without any approximation for
$N=1,000$, the black line refers to the first-order approximate expression
(\ref{eq:lspecb3}) and the dashed blue line to the approximation
(\ref{eq:specb2}). In the inset we magnify the region of small values of
$\f$.
The parameters are $a= 1.3$, $g=-1.2$, and $\beta=1.0$.}
\label{fig5}
\end{figure}

\subsection{Phase diagram}

The main difference with respect to the case of finite pulse--width
is that the spectral component associated with SWs
does not reduce to a fully degenerate zero Floquet exponent. On the contrary, 
it actively contributes to determining the stability properties of the network. 
While the splay state is found to be unstable for finite values of $\beta$ and 
excitatory coupling ($g > 0$), a more interesting scenario is found for the 
inhibitory case ($g < 0$)\, . 

Let us illustrate such situations by analyzing the stability spectrum obtained
for $a=1.3$, $\beta=1.0$ and $g=-1.2$. 
In Fig.~\ref{fig5} we show that the theoretical
expression~\eqref{eq:lspecb3} reproduces most of the spectrum obtained by the
numerical diagonalization of the exact Jacobian for a network of 1000 neurons.
In particular, the agreement is very good around the top part of the Floquet
spectrum, which accounts for the stability of the network. More precisely, the
least stable mode is that one characterized by the highest frequency
(the $\pi$-mode), i.e. it corresponds to an up-down behavior of the
perturbation, when passing from one to the next neuron. As this condition
holds true for arbitrary values of $N$, we are led to conclude that one 
cannot perfom the continuum limit straightforwardly,
since this would remove those fluctuations that are most
relevant for the stability of the network. 

Notice also that the only region where Eq.~\eqref{eq:lspecb3} fails to reproduce 
the true spectrum is close to vanishing values of the frequency $\f$. 
Analogously to the previous case, in this region (see the inset in Fig.~\ref{fig5})
one has to invoke Eq.~(\ref{eq:specb2}) to account for the dip centered around
$\f \sim 0$.
\begin{figure}[t!]
\includegraphics[draft=false,clip=true,height=0.34\textwidth]{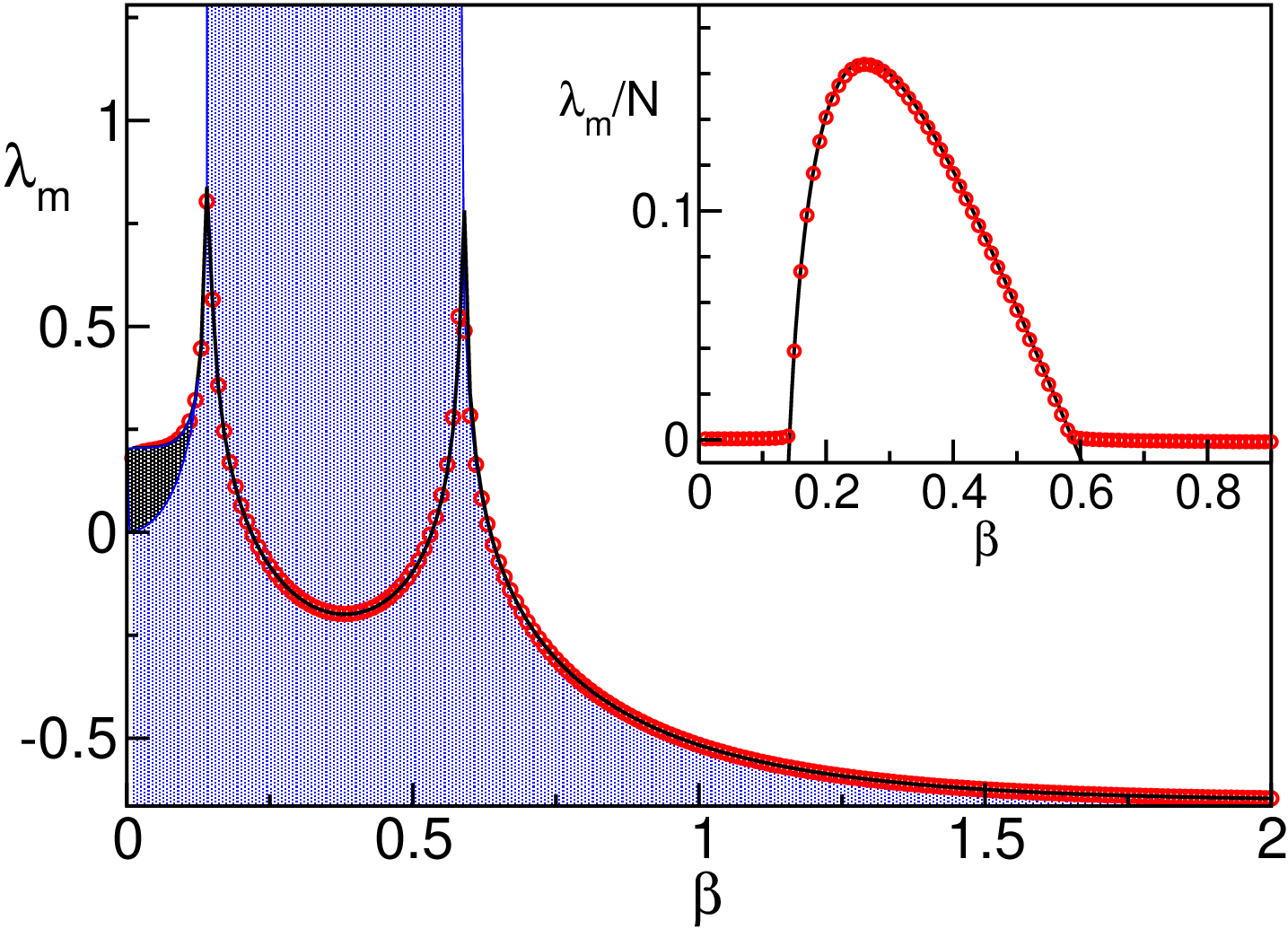}
\caption{(Color online) Maximal Floquet exponent $\lambda_m$ as a function of $\beta$ for
$a=1.3$ and $g=-1.2$. The line bordering the shaded region corresponds to the
numerically obtained maximum in a network of $N=500$ neurons; red circles
correspond to the second eigenvalue in the same network, while the solid line
corresponds to the analytical expression (\ref{eq:lanab}) for $\lambda_{\pi}$. 
In particular, the lower border of the dark shaded region corresponds to $\lambda_{\pi}$. In the inset,
the numerically computed Floquet exponent (open red circles) is compared to
the analytical expression (\ref{eq:mubeta}) for $\lambda_{N,N+1}$.}
\label{fig6}
\end{figure}

If one has to determine the entire spectrum of a finite network, the 
two components arising from Eqs.~(\ref{eq:specb2},\ref{eq:lspecb3}) have
to be properly matched. After selecting equispaced modes according to
the system size (the spacing being $2\pi/N$), one has to identify the
value of $k$ where the distance between the two spectral components is minimal.
Below this value the component to be considered is that of LWs
(see Eq.~(\ref{eq:specb2})), while above such a value the spectrum
is well approximated by Eq.~\eqref{eq:lspecb3}\, .

It is also instructive to investigate the dependence of the spectrum
on the parameter $\beta$. In Fig.~\ref{fig6} we have plotted the maximum Floquet
exponent $\lambda_m$ as a function of $\beta$. The maximum has been determined by
diagonalizing numerically the Jacobian for a network of $N=500$ neurons: it
corresponds to the border of the shaded region (for the sake of clarity, the
peak has been cut out). The comparison with the results obtained
from Eq.~(\ref{eq:lspecb3}) confirms that SW modes account
for the stability of the entire network in a quite wide range of $\beta$ values,
except for an intermediate region, where the maximum Floquet exponent is well 
reproduced by Eq.~\eqref{eq:mubeta} (as it can be appreciated by looking at
the inset of Fig.~\ref{fig6}). 

\begin{figure}[t!]
\includegraphics[draft=false,clip=true,height=0.34\textwidth]{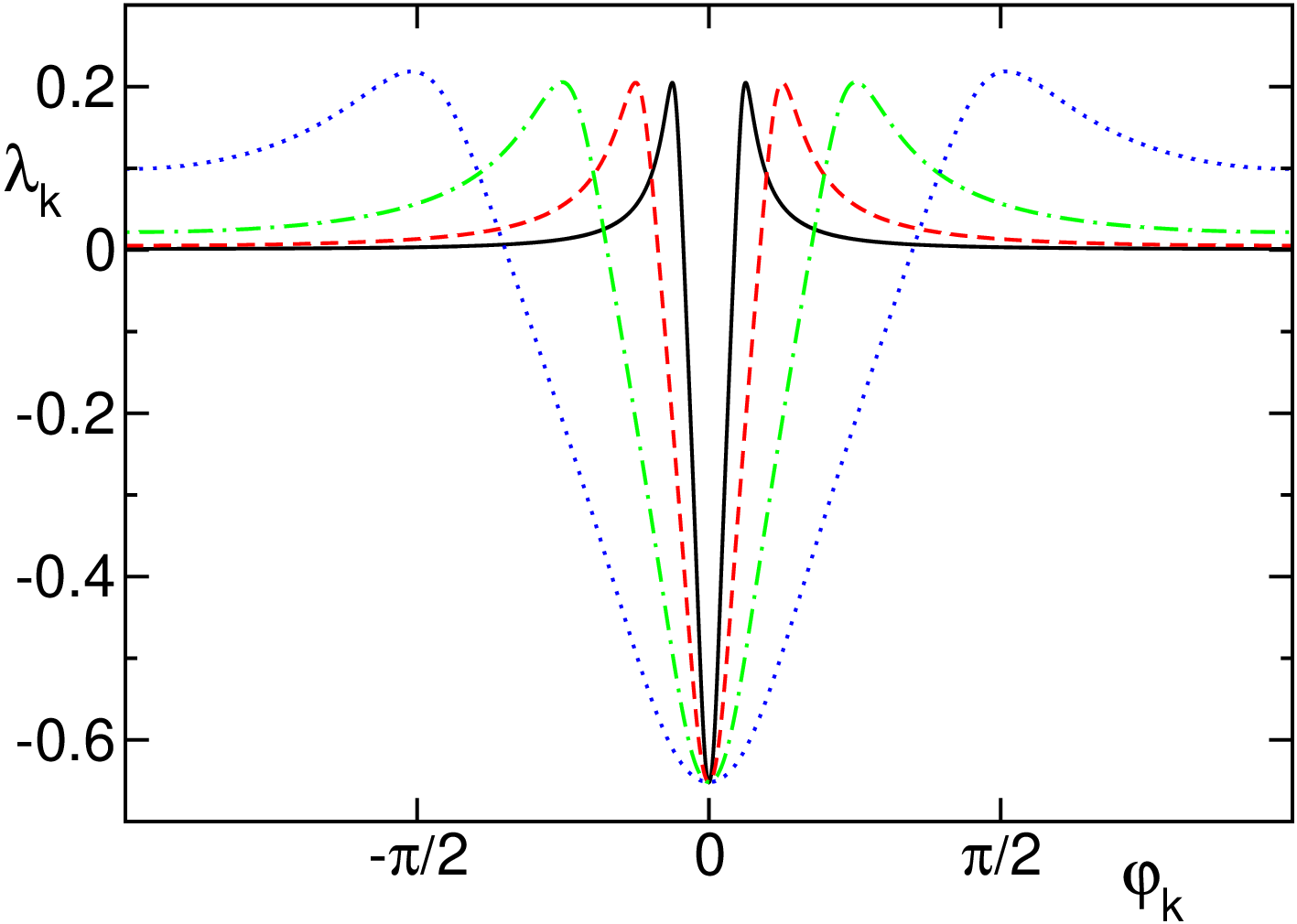}
\caption{(Color online) Finite-frequency Floquet spectra $\lambda(\varphi)$ for $\beta=0.01$ (black solid line), 0.02 (red dashed line), 
0.04 (green dot-dashed line) and 0.08 (blue dotted line) and $a=1.3$, $g=-1.2$.}
\label{fig7}
\end{figure}

Outside this region, $\lambda_m$ is well reproduced by the analytical
expression Eq.~(\ref{eq:lanab}), except for $\beta < 0.125$. Indeed, for small 
$\beta$, the maximum of the spectrum occurs for $|\varphi| <\pi$, as it can be 
seen in Fig.~\ref{fig7}. Moreover, the maximal Floquet exponent $\lambda_{m}$ 
remains finite even for $\beta \to 0$. Finally, inside the intermediate region 
dominated by $\lambda_{N,N+1}$, the analytical prediction (\ref{eq:lanab}) is
extremely close to the 2nd (numerically computed) exponent. Therefore, the theory 
exposed in this section predicts correctly the stability of the network, although
one must carefully identify the spectral component that is responsible for the 
relevant contribution.

There is another important conclusion that can be drawn from Fig.~\ref{fig6},
by comparing two limit cases. On the one hand, the limit $\beta \to \infty$
corresponds to $\delta$-like pulse. In fact, one can see that the maximum
Floquet exponent coincides with the expression derived in Ref.~\cite{zillmer}.
On the other hand, for $\beta=0$, the model converges to
the $\alpha \to \infty$ limit of the model considered in \cite{abbott}. 
In the two limits, 
\begin{equation}\label{eq:lanab0}
  \lambda_{\pi}=-\frac{g \beta^2T^2}{6[(a-1)T+g](aT+g)}
\end{equation}

Altogether, the $\beta$-dependence resolves the contradiction mentioned in the
beginning of this section, as we can conclude that the pulse--width does not only
affect the quantitative value of the Floquet exponent, but contributes also
to determining the qualitative stability properties. It is therefore useful
to think about the meaning of $\beta$. It expresses the pulse bandwidth
$1/\alpha$ in terms of ``$1/N$" units. By remembering that the interspike
interval is $T/N$, it is convenient to introduce the adimensional
parameter
$$
 r = \frac{T/N}{1/\alpha} = \beta T  \, .
$$
This parameter, being the ratio between the interspike interval and the
pulse--width, is susceptible of being determined in general contexts that go
beyond the specific pulse-shape choice and model adopted in the
present paper.

\begin{figure}[t!]
\includegraphics[draft=false,clip=true,height=0.34\textwidth]{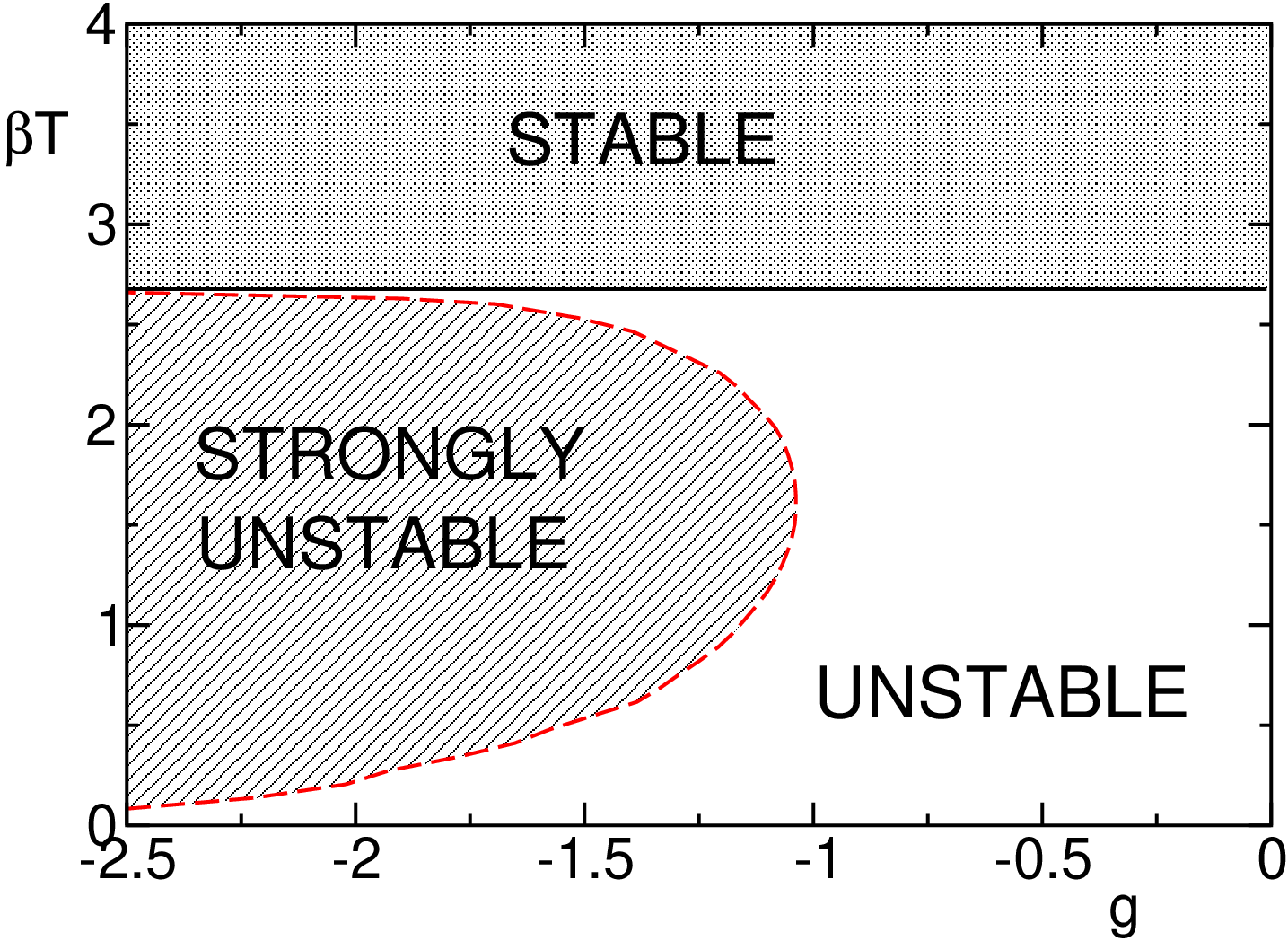}
\caption{(color online) Phase diagram for the stability of the splay state for $a=1.3$ and
inhibitory coupling in the sharp spike limit. The solid black horizontal
line divides the stable from the unstable region, while the dashed red line encircle the strongly
unstable phase where the maximum Floquet exponent is proportional to $N$.}
\label{fig8}
\end{figure}

Actually, the phase diagram reported in Fig.~\ref{fig8}, confirms that
$r=\beta T$ is a meaningful parameter, since stable and unstable phases are
separated by a critical line, where $r$ is constant independently of $g$.
By solving $\lambda_{\pi} = 0$, we find that the critical value of $r$
discriminating between stable and unstable phase is given by the following
implicit equation
\begin{equation}\label{eq:critical_r}
  {\rm e}^{4r} -2(r^2+1){\rm e}^{3r} -2 r^2 {\rm e}^r +1 = 0 \quad .
\end{equation}
Moreover, we see that the strongly unstable regime seen in Fig.~\ref{fig6} arises
only for a sufficiently strong inhibitory coupling ($g < -1$).

\begin{figure}[t!]
\includegraphics[draft=false,clip=true,height=0.34\textwidth]{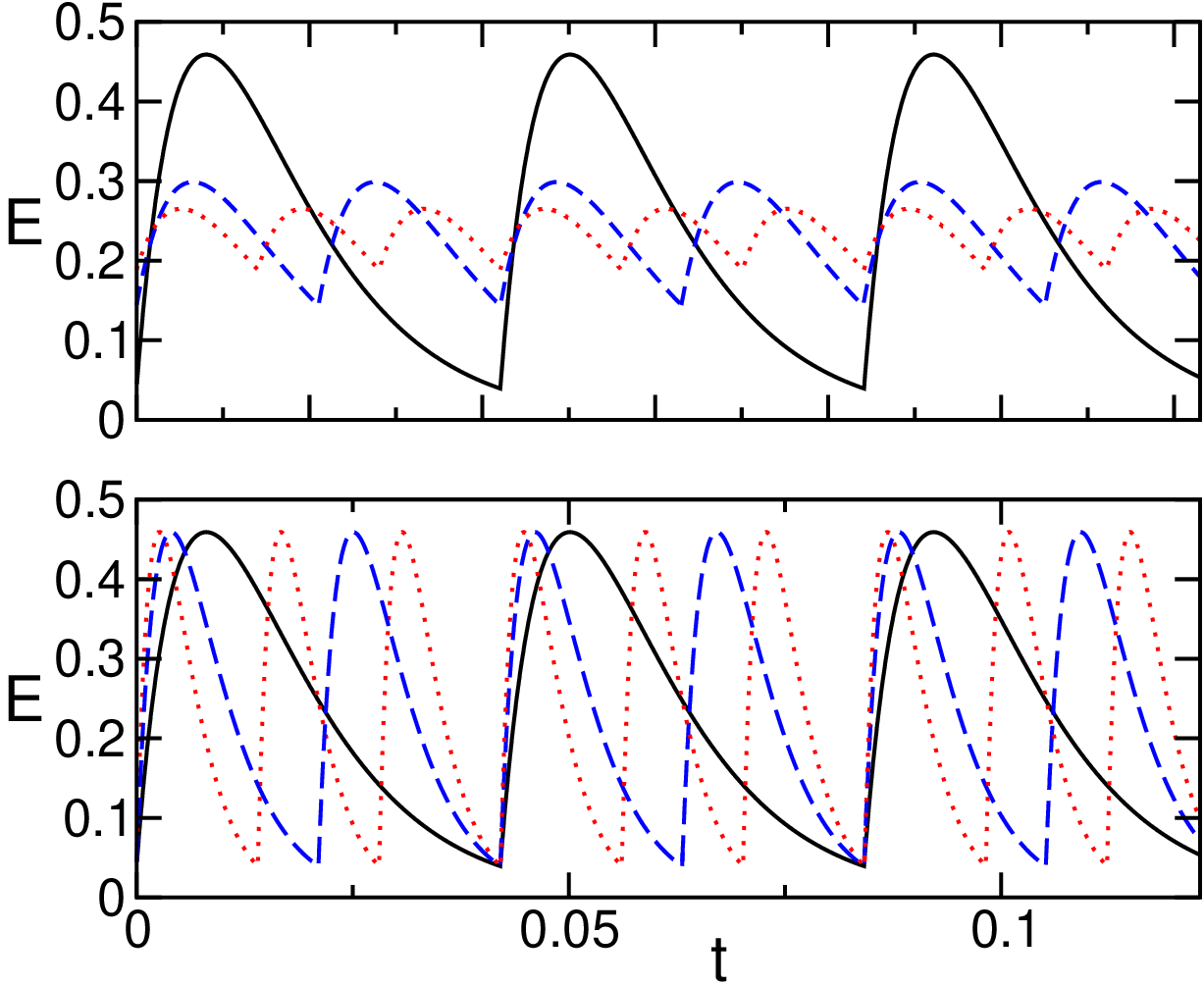}
\caption{(Color online) Time evolution of the field $E(t)$ associated to a splay state
for increasing $N$ values: $N=100$ (black solid lines), 200 (blue dashed lines), 300 (red dotted lines).
The upper panel refers to a fixed value of $\alpha= 120$, while the lower panel to
a constant $\beta=0.14$. The other parameters are $a=1.3$ and $g=-1.2$}
\label{fig9}
\end{figure}

We conclude this section by coming back to the possible reasons for the failure
of the mean field approach. In the upper panel of Fig.~\ref{fig9}, we show
the time evolution of the field $E(t)$ for increasing values of $N$
and fixed $\alpha=120$. The oscillations around the mean value tend to decrease
and this indeed suggests that it is meaningful to introduce a sort of average
flux of spiking neurons as done in the mean field approach. One can also see
that since $E(t)$ is increasingly flat (for increasing $N$), it is natural
to expect that the stability of the splay state is very weak: The neuron-neuron
interactions are in fact mediated by the common field $E$. It might be tempting
to reduce the stability of the splay state to that of the single neuron exposed
to a given field dynamics $E(t)$, but this would amount to no more than a crude
approximation.
In other words, there is no shortcut for performing the analysis carried out
in this paper.

In the lower panel, of Fig.~\ref{fig9}, the field evolution is reported for
the same values of $N$ and $\beta=0.14$. In this case, field oscillations
become faster but their size does not decrease upon increasing $N$. We consider
this as the major reason for the failure of the mean field approach.
Furthermore the finite amplitude of the oscillations is also responsible for
mantaining a finite stability even in the limit $N\to\infty$\, .

\section{Conclusions and perspectives}

In this paper we have shown that the stability of splay states can be addressed
by reducing a model of globally coupled differential equations to suitable event-driven
maps which relate the internal configuration at two consecutive spike-emissions.
The analytical investigation of the Jacobian in the large $N$ limit reveals that 
the spectrum of eigenvalues is made of three components: {\it i)} long wavelengths
eigenmodes that emerge also from a mean-field approach \cite{abbott}; {\it ii)} 
short wavelengths; {\it iii)} isolated eigenvalues, which signal the existence 
of strong instabilities, i.e. eigenvalues that are proportional to the network size.
Altogether, we find that drastically different results can be found for different 
values of the ratio $r$ between the interspike interval and the pulse--width.
This leads us to conclude that the stability of large networks of neurons
coupled via narrow pulses, crucially depends on the parameter $r$, 
thus suggesting that the dynamycal stability of these models demands a 
more refined treatment than mean--field. It will be interesting to 
investigate the role of $r$ in more general contexts,
e.g., by considering different pulse shapes (possibly adding the effect of
delay) and different force fields (possibly including further degrees of
freedom, as it naturally appears in the Hodgkin-Huxley model).

Another issue that it will be worth analyzing concerns the exact stability
in large but finite networks in the presence of finite pulse-widths.
Our analysis has revealed that the first order approximations fails to 
reproduce even the sign of the maximum Floquet exponent. That is a consequence 
of the $1/N^2$ decay of the spectrum which, in turn, requires developing a 
third-order perturbation theory. Our numerics shows that the splay state is always 
stable towards SW perturbations in networks of LIF neurons with 
excitatory coupling, while a 1st and 2nd order approximation theories may give 
rise to unstable states. Is this an indication of a systematic drawback of  
discrete-time models, or an indication that qualitatively different stability 
properties may be found for different force fields? The analysis of nonlinear
force fields is definitely needed for obtaining a convincing answer to this question.


\end{document}